\documentstyle[12pt]{article}
\begin{document}
{\pagestyle{empty}
\rightline{TEZU-F-087}
\rightline{November 1994}
\vskip 3cm
\centerline{\large \bf On Solutions of Tetrahedron Equations}
\centerline{\large \bf based on Korepanov Mechanism}
\vskip 2cm
\centerline{Minoru Horibe \footnote{E-mail address:
horibe@newton.apphy.fukui-u.ac.jp}}
\centerline {{\it Department of Physics, Faculty of Education}}
\centerline {{\it Fukui University,Fukui 910, Japan}}
\vskip 0.3cm
\centerline{Kazuyasu Shigemoto \footnote{E-mail address:
shigemot@tezukayama-u.ac.jp}}
\centerline {{\it Department of Physics}}
\centerline {{\it Tezukayama University, Nara 631, Japan }}
\vskip 2cm
\centerline{\bf Abstract} \vspace{10mm}
%\vskip 0.2in
We have examined solutions of tetrahedron equations from the elliptic free
fermion model by using Korepanov mechanism based on tetrahedral
Zamolodchikov algebras.  As a byproduct, we have found a new integrable
2-dim. lattice model.  We have also studied the relation between
tetrahedral Zamolodchikov algebras and tetrahedron equations.
\newpage}
%%%%%%%%%%%%%%%%%%%%%%% Section 1 %%%%%%%%%%%%%%%%%%%%%%%%
\noindent {\bf \S 1.\ \ Introduction} \vspace{2mm}

\indent
In last two decades, much understanding has been made in 2-dim.
integrable statistical models with beautiful mathematical
structures~\cite{qg},  and recently much attention has
turned to 3-dim. integrable statistical models.

Tetrahedron equations are 3-dim. generalizations of Yang-Baxter
equations in 2-dim., and Zamolodchikov first found solutions
of tetrahedron equations~\cite{Zamolodchikov},
and Baxter confirmed Zamolodchikov's result~\cite{Baxter2}.
After these pioneering works, little progress~\cite{B-S}~\cite{Maillard}
has been made until recently because of difficulties of the problem.
After Bazhanov-Baxter~\cite{B-B}
have obtained solutions of tetrahedron equations for $N$\ colors model,
by generalizing Zamolodchikov's solutions of $2$\ colors model, many
interesting papers have published on solutions of
tetrahedron equations
{}~\cite{Mangazeev,Saito,Hietarinta,Korepanov1,Korepanov2,Korepanov3}.

In this paper, we take Korepanov's approach and examine solutions
of tetrahedron equations from the elliptic free fermion model.
In section 2, we review Korepanov's approach and examine the relation
between tetrahedral Zamolodchikov algebras and tetrahedron equations.
In section 3, we give explicit solutions of tetrahedron equations
from the elliptic free fermion model. \vspace{10mm}

%\hfil\break\indent
%%%%%%%%%%%%%%%%%%%%%%%%%%% section 2 %%%%%%%%%%%%%%%%%%%%%%
\noindent {\bf \S 2. Relation between Tetrahedral Zamolodchikov Algebras
and Tetrahedron Equations \quad -\ Korepanov Mechanism\ -} \vspace{2mm}

\indent
We follow the notation of Korepanov, and
denote the standard $R$-matrices as $R^0$ and the related
non-symmetric $R$-matrices as $R^1$~\cite{Korepanov1,Korepanov2,Korepanov3}.
For the 8-vertex case, $R^0$\ and $R^1$\ are written in the forms
\begin{eqnarray}
R^0_{ij}(u_i,u_j)=f_0(u_i,u_j)
\left(\begin{array}{cccc}
a_{ij} & 0 & 0 & d_{ij} \\
0 & b_{ij} & c_{ij} & 0 \\
0 & c_{ij} & b_{ij} & 0 \\
d_{ij} & 0 & 0 & a_{ij}     \end{array} \right)\ ,
\label{e1}
\end{eqnarray}
\begin{eqnarray}
R^1_{ij}(u_i,u_j)=f_1(u_i,u_j)
\left(\begin{array}{cccc}
-a^{'}_{ij} & 0 & 0 & d^{'}_{ij} \\
0 & -b^{'}_{ij} & c^{'}_{ij} & 0 \\
0 & -c^{'}_{ij} & b^{'}_{ij} & 0 \\
-d^{'}_{ij} & 0 & 0 & a^{'}_{ij}     \end{array} \right)\ ,
\label{e2}
\end{eqnarray}
where
\begin{eqnarray}
&&a_{ij}={\rm sn}(\lambda-u_i+u_j) \ , \quad
b_{ij}={\rm sn}(u_i-u_j)\ ,  \nonumber \\
&&c_{ij}={\rm sn} (\lambda)\ , \qquad \qquad \quad
d_{ij}=k {\rm sn}(\lambda) {\rm sn}(u_i-u_j)
{\rm sn}(\lambda-u_i+u_j)\ ,
\label{e3}
\end{eqnarray}
\begin{eqnarray}
&&a^{'}_{ij}={\rm sn}(\lambda-u_i-u_j)\ ,  \quad
b^{'}_{ij}={\rm sn}(u_i+u_j)\ ,  \nonumber \\
&&c^{'}_{ij}={\rm sn}(\lambda)\ , \qquad \qquad \quad
d^{'}_{ij}=k {\rm sn}(\lambda) {\rm sn}(u_i+u_j)
{\rm sn}(\lambda-u_i-u_j)\ ,
\label{e4}
\end{eqnarray}
where ${\rm sn}(u)$ is one of the Jacobi's elliptic functions, $k$ is the
modulus of the elliptic function, and $\lambda$\ is an arbitrary constant
parameter.  The functions $f_0(u_i,u_j)$ and $f_1(u_i,u_j)$ are arbitrary
here, and we choose the special functional forms in section 3.  We may
use short-hand notations $R^a_{ij}$ instead of full
notations $R^a_{ij}(u_i,u_j)$.

Using the above parametrization, $R$-matrices satisfy the
following Yang-Baxter equations
\begin{eqnarray}
&&R^0_{12}(u_1,u_2) R^0_{13}(u_1,u_3) R^0_{23}(u_2,u_3) \nonumber \\
&&= R^0_{23}(u_2,u_3) R^0_{13}(u_1,u_3) R^0_{12}(u_1,u_2) \ ,
\label{e5}
\end{eqnarray}
where $R^{a}_{ij}$\ act as ${\rm End}(V^{\otimes 3})$. Operatores
$R^{a}_{12}$,
for example,  acts on $V_1 \otimes V_2$\ as non-trivial $R$-matrices
and on $V_3$\ as identity.

First, we have found that $R^0_{ij}$\ and $R^1_{ij}$\ satisty the
following equations
\begin{eqnarray}
&&R^0_{12}(u_1,u_2) R^1_{13}(u_1,u_3) R^1_{23}(u_2,u_3) \nonumber \\
&&= R^1_{23}(u_2,u_3) R^1_{13}(u_1,u_3) R^0_{12}(u_1,u_2) \ ,
\label{e6}
\end{eqnarray}
which reflect the symmetry of $R^0_{ij}$\ and $R^1_{ij}$.
Eq.(\ref{e6}) gives us a new example of integrable 2-dim. lattice
models with Boltzmann weights $R^1_{ij}$.
In order that transfer matrices which are composed of $R^1_{ij}$
commute, it is sufficient to
exist the invertible $R^0_{ij}$  which satisfy
$R^0_{12} R^1_{13} R^1_{23}=R^1_{23} R^1_{13} R^0_{12}$.
Thus Eq.(\ref{e6}) gives us a new class of integrable 2-dim.
lattice models, though its Boltzmann
weights do not satisfy the physical
positivity condition.  Though this new
integrable model is unphysical,
it will be helpful to study new mathematics behind this type of
integrable models.

Next we consider Korepanov mechanism to construct solutions of
tetrahedron equations (Zamolodchikov equations).
Korepanov consider the following tetrahedral Zamolodchikov algebras,
which represent {\it scattering} relations
\begin{eqnarray}
&&R^a_{12}(u_1,u_2) R^b_{13}(u_1,u_3) R^c_{23}(u_2,u_3)\nonumber \\
&&=\sum_{d,e,f=0}^{1} S(u_1,u_2,u_3)^{abc}_{def} \
R^f_{23}(u_2,u_3) R^e_{13}(u_1,u_3) R^d_{12} (u_1,u_2) \ ,
\label{e7}
\end{eqnarray}
where $S^{abc}_{def}$\ are the $S$-{\it matrices} of
{\it particles} $R^a_{ij}$.
Here Korepanov have
taken the very special forms, that is,
2 dim. integrable $R$-matrices $R^0_{ij}$ and their
related $R^1_{ij}$ matrices, for {\it particles}.

Tetrahedral Zamolodchikov algebras (\ref{e7}) are
overdeterministic in a sense
that, there are 64 equations with only 8
variables $S^{abc}_{def}$ for
fixed $\{a,b,c \}$ in general, because $R^a_{12}R^b_{13}R^c_{23}$\ and
$R^f_{23}R^e_{13}R^d_{12}$ are $8 \times 8$\ matrices, while
$S^{abc}_{def}$\ are simple numberes for given $\{a,b,c,d,e,f\}$.
Because of the special symmetry of these $R^0_{ij}$\ and $R^1_{ij}$,
only 20 equations are not automatically satisfied, while the number of
the variables $S^{abc}_{def}$\ to be solved is  8 for the fixed $\{a,b,c\}$.
Tetrahedral Zamolodchikov algebras and variables
$S^{abc}_{def}$\ can be decomposed
into even and odd sectors.  Even and odd variables are
those of $i)\ a+b+c \equiv d+e+f\ ({\rm mod}\ 2)$\ and
$ii)\ a+b+c \equiv d+e+f+1\ ({\rm mod}\ 2)$
cases in $S^{abc}_{def}$ respectively.
Odd sector can be easily solved to give the trivial
solution $S^{abc}_{edf}=0\ (a+b+c \equiv d+e+f+1\ ({\rm mod}\ 2))$.
Thus non-trivial equations to be solved are only in even
sector, and the number of equations to be solved becomes only 10, while the
number of variables $S^{abc}_{def}$ becomes only 4 for fixed $\{a,b,c \}$
but still overdeterministic. We will discuss on this
overdeterministic problem in section 3 in the process of solving
tetrahedral Zamolodchikov algebras.

If there exist solutions $S^{abc}_{def}$\ for
given $R^0_{ij}$ and $R^1_{ij}$ in tetrahedral Zamolodchikov algebras,
it is quite promising that these $S^{abc}_{def}$\ satisfy
the following tetrahedron equations
\begin{eqnarray}
\sum_{a',b',c', \atop d',e',f'}
&&S(u_{1},u_{2},u_{3})^{abc}_{a'b'c'} \
   S(u_{1},u_{2},u_{4})^{a'de}_{a''d'e'} \nonumber \\
&& \times  S(u_{1},u_{3},u_{4})^{b'd'f}_{b''d''f'} \
   S(u_{2},u_{3},u_{4})^{c'e'f'}_{c''e''f''} \nonumber\\
=\sum_{a',b',c', \atop d',e',f'}
&&S(u_{2},u_{3},u_{4})^{cef}_{c'e'f'} \
   S(u_{1},u_{3},u_{4})^{bdf'}_{b'd'f''} \nonumber \\
&& \times   S(u_{1},u_{2},u_{4})^{ad'e'}_{a'd''e''} \
   S(u_{1},u_{2},u_{3})^{a'b'c'}_{a''b''c''} \ .
\label{e8}
\end{eqnarray}
This mechanism to find solutions of tetrahedron
equations by using solutions of tetrahedral Zamolodchikov
algebras, we call Korepanov mechanism.

We denote the above equations in the following short-hand way;
\begin{eqnarray}
& &S_{123}(u_{1},u_{2},u_{3}) \ S_{145}(u_{1},u_{2},u_{4}) \
   S_{246}(u_{1},u_{3},u_{4}) \ S_{356}(u_{2},u_{3},u_{4}) \nonumber\\
&=&S_{356}(u_{2},u_{3},u_{4}) \ S_{246}(u_{1},u_{3},u_{4}) \
   S_{145}(u_{1},u_{2},u_{4}) \ S_{123}(u_{1},u_{2},u_{3}) \ ,
\label{e9}
\end{eqnarray}
where $S^{abc}_{def}$ act as ${\rm End}(V^{\otimes 6})$.
Operators $S_{123}$, for example,
acts on $V_1 \otimes V_2 \otimes V_3$ as
$S^{abc}_{def}$ and on the rest as identity.

Here we sketch the reason why solutions of
tetrahedral Zamolodchikov algebras
are promising candidates to solutions of tetrahedron equations.
We start from the following objects
\begin{eqnarray}
 R^{a_1}_{12} R^{a_2}_{13} R^{a_3}_{23}
 R^{a_4}_{14} R^{a_5}_{24} R^{a_6}_{34} \ ,
\label{e11}
\end{eqnarray}
and we successively use tetrahedral Zamolodchikov
algebras (\ref{e7}) for
$R^{a_1}_{12} R^{a_2}_{13} R^{a_3}_{23}$,
$R^{b_1}_{12} R^{a_4}_{14} R^{a_5}_{24}$,
$R^{b_2}_{13} R^{c_4}_{14} R^{a_6}_{34}$,
$R^{b_3}_{23} R^{c_5}_{24} R^{d_6}_{34}$, starting from the above object
(\ref{e11}). We then obtain
\begin{eqnarray}
 R^{a_1}_{12} R^{a_2}_{13} R^{a_3}_{23}
R^{a_4}_{14}  R^{a_5}_{24} R^{a_6}_{34}
\nonumber \\
=\sum_{b_1,b_2,b_3,b_4,b_5,b_6, \atop c_1,c_2,c_3,c_4,c_5,c_6}&&
 S(u_1,u_2,u_3)^{a_1a_2a_3}_{b_1b_2b_3}\
 S(u_1,u_2,u_4)^{b_1a_4a_5}_{c_1b_4b_5}\
 S(u_1,u_3,u_4)^{b_2b_4a_6}_{c_2c_4b_6}  \nonumber \\
&& \times S(u_2,u_3,u_4)^{b_3b_5b_6}_{c_3c_5c_6}\
 R^{c_6}_{34} R^{c_5}_{24} R^{c_4}_{14}
 R^{c_3}_{23} R^{c_2}_{13} R^{c_1}_{12} \ .
\label{e12}
\end{eqnarray}
Here we have used the fact that $[R^a_{ij},R^b_{kl}]=0$ for
$i \ne k,l$ and $j \ne k,l$, for example
$[R^a_{12},R^b_{34}]=0$, from the property of the tensor product.

Next, starting from the same objects (\ref{e11}), we successively
use tetrahedral Zamolodchikov algebras for
$R^{a_3}_{23} R^{a_5}_{24} R^{a_6}_{34}$,
$R^{a_2}_{13} R^{a_4}_{14} R^{c_6}_{34}$,
$R^{a_1}_{12} R^{d_4}_{14} R^{c_5}_{24}$,
$R^{c_1}_{12} R^{d_2}_{13} R^{c_3}_{23}$,
which is different from the previous ordering in
the use of tetrahedral Zamolodchikov algebras.
In this ordering, we obtain
\begin{eqnarray}
 R^{a_1}_{12} R^{a_2}_{13} R^{a_3}_{23}
R^{a_4}_{14}  R^{a_5}_{24} R^{a_6}_{34}
\nonumber \\
=\sum_{b_1,b_2,b_3,b_4,b_5,b_6,\atop c_1,c_2,c_3,c_4,c_5,c_6}&&
 S(u_2,u_3,u_4)^{a_3a_5a_6}_{b_3b_5b_6} \
 S(u_1,u_3,u_4)^{a_2a_4b_6}_{b_2b_4c_6} \
 S(u_1,u_2,u_4)^{a_1b_4b_5}_{b_1c_4c_5} \nonumber \\
&& \times  S^{b_1b_2b_3}_{c_1c_2c_3}(u_1,u_2,u_3) \
 R^{c_6}_{34} R^{c_5}_{24} R^{c_4}_{14}
 R^{c_3}_{23} R^{c_2}_{13} R^{c_1}_{12} \ .
\label{e13}
\end{eqnarray} From Eqs.(\ref{e12}) and (\ref{e13}), we
obtain
\begin{eqnarray}
\sum_{b_1,b_2,b_3,b_4,b_5,b_6, \atop c_1,c_2,c_3,c_4,c_5,c_6}&&
 S(u_1,u_2,u_3)^{a_1a_2a_3}_{b_1b_2b_3} \
 S(u_1,u_2,u_4)^{b_1a_4a_5}_{c_1b_4b_5} \
 S(u_1,u_3,u_4)^{b_2b_4a_6}_{c_2c_4b_6} \nonumber \\
&& \times S(u_2,u_3,u_4)^{b_3b_5b_6}_{c_3c_5c_6}  \
 R^{c_6}_{34} R^{c_5}_{24} R^{c_4}_{14}
 R^{c_3}_{23} R^{c_2}_{13} R^{c_1}_{12}
\nonumber \\
=\sum_{b_1,b_2,b_3,b_4,b_5,b_6, \atop c_1,c_2,c_3,c_4,c_5,c_6}&&
 S(u_2,u_3,u_4)^{a_3a_5a_6}_{b_3b_5b_6} \
 S(u_1,u_3,u_4)^{a_2a_4b_6}_{b_2b_4c_6} \
 S(u_1,u_2,u_4)^{a_1b_4b_5}_{b_1c_4c_5} \nonumber \\
&&\times  S(u_1,u_2,u_3)^{b_1b_2b_3}_{c_1c_2c_3} \
 R^{c_6}_{34} R^{c_5}_{24} R^{c_4}_{14}
 R^{c_3}_{23} R^{c_2}_{13} R^{c_1}_{12} \ .
\label{e14}
\end{eqnarray}
These relations strongly suggest that the following Zamolodchikov's
tetrahedron equations are satisfied
\begin{eqnarray}
\sum_{b_1,b_2,b_3, \atop b_4,b_5,b_6}&&
 S(u_1,u_2,u_3)^{a_1a_2a_3}_{b_1b_2b_3}  \
 S(u_1,u_2,u_4)^{b_1a_4a_5}_{c_1b_4b_5} \nonumber \\
&& \times
 S(u_1,u_3,u_4)^{b_2b_4a_6}_{c_2c_4b_6} \
 S(u_2,u_3,u_4)^{b_3b_5b_6}_{c_3c_5c_6}  \nonumber \\
=\sum_{b_1,b_2,b_3, \atop b_4,b_5,b_6}&&
 S(u_2,u_3,u_4)^{a_3a_5a_6}_{b_3b_5b_6} \
 S(u_1,u_3,u_4)^{a_2a_4b_6}_{b_2b_4c_6} \nonumber \\
&& \times
 S(u_1,u_2,u_4)^{a_1b_4b_5}_{b_1c_4c_5} \
 S(u_1,u_2,u_3)^{b_1b_2b_3}_{c_1c_2c_3} \ ,
\label{e15}
\end{eqnarray}
which are equaivalent to short-hand forms (\ref{e9}).

In this way, if tetrahedron equations (\ref{e15}) are satisfied,
integrability conditions (\ref{e14}) of tetrahedral
Zamolodchikov algebras are automaitcally satisfied, but the opposite
is not always true.
According to Korepanov mechanism, we must first solve
tetrahedral Zamolodchikov algebras to find $S^{abc}_{def}$,
which are promising candidates
to satisfy tetrahedron equations.
Next we must check explicitly whehter these $S^{abc}_{def}$ really satisfy
tetrahedron equations or not. \vspace{10mm}

%%%%%%%%%%%%%%%%%%%%%%%%%%% section 3 %%%%%%%%%%%%%%%%%%%%%%
\noindent {\bf \S 3. Solutions of Tetrahedron Equations from
Elliptic Free Fermion Model}\vspace{2mm}

\indent
Here we start to solve tetrahedral Zamolodchikov algebras.
We consider first the sector $\{a=0,b=0,c=0 \}$\ in
tetrahedral Zamolodchikov algebras (\ref{e7}). From Yang-Baxter
equations (\ref{e5}), we can see that
$S(u_{1},u_{2},u_{3})^{000}_{def}=\delta_{d,0}\delta_{e,0}\delta_{f,0}$
are one of the solutions. In the eight vertex case,
solutions become unique, and the above are only unique solutions.
The same situation happens for the sector $\{a=0,b=1,c=1 \}$, and
solutions are uniquely determined to give
$S(u_{1},u_{2},u_{3})^{011}_{def}=\delta_{d,0}\delta_{e,1}\delta_{f,1}$.
For $\{a=1,b=0,c=1 \}$ and $\{a=1,b=1,c=0 \}$ sectors, some
delicate thing happens. From four equations  we obtain solutions
$S(u_{1},u_{2},u_{3})^{101}_{def}=\delta_{d,1}\delta_{e,0}
\delta_{f,1}$ and $S(u_{1},u_{2},u_{3})^{110}_{def}
=\delta_{d,1}\delta_{e,1}\delta_{f,0}$ for $\{a=1,b=0,c=1 \}$ and
$\{a=1,b=1,c=0 \}$ sectors respectively.
But because of the overdeterministic property of
tetrahedral Zamolodchikov algebras,
there are still six equations to be satisfied, and
we can see that these six equations  are not satisfied in this 8 vertex
parametrization, that is, we can conclude that there are no solutions
for tetrahedral Zamolodchikov algebras in the 8 vertex parametrization.
Then we must consider more restricted cases to find solutions
of tetrahedral Zamolodchikov algebras.   One of the cases which give
solutions of tetrahedral Zamolodchikov algebras is the
symmetric free fermion case.
The symmetric free fermion case is the
special case of the 8 vertex case, that is, it is the case that
$\lambda$\ in Eqs.(\ref{e3}) and (\ref{e4}) becomes the complete
elliptic integral
$\displaystyle{ \lambda=K=\int_{0}^{1}
\frac{dx}{ \sqrt{(1-x^2)(1-k^2 x^2)} }  }$.
In this special case, we can easily check that the free fermion
condition $a^2+b^2=c^2+d^2$\ is satisfied~\cite{Felderhof,B-S2}.
Choosing $\displaystyle{ f_0(u_i,u_j)= \frac{ {\rm dn}(u_i-u_j)}
{ {\rm sn}(u_i-u_j) }  }$,
and $\displaystyle{  f_1(u_i,u_j)=\frac{1}{ {\rm cn}(u_i+u_j) } }$,
$R^0_{ij}$ and $R^1_{ij}$ become in the forms
\begin{eqnarray}
R^0_{ij}(u_i,u_j)=
\left(\begin{array}{cccc}
\displaystyle{\frac{{\rm cn}(u_i-u_j)}{{\rm sn}(u_i-u_j)}} & 0 &
0 & k {\rm cn}(u_i-u_j) \\
0 & {\rm dn}(u_i-u_j) &
\displaystyle{\frac{{\rm dn}(u_i-u_j)}{{\rm sn}(u_i-u_j)}} & 0 \\
0 & \displaystyle{\frac{{\rm dn}(u_i-u_j)}{{\rm sn}(u_i-u_j)}} &
{\rm dn}(u_i-u_j) & 0 \\
k {\rm cn}(u_i-u_j) & 0 &
0 & \displaystyle{\frac{{\rm cn}(u_i-u_j)}{{\rm sn}(u_i-u_j)}}
\end{array} \right) \ ,
\label{e16}
\end{eqnarray}
\begin{eqnarray}
R^1_{ij}(u_i,u_j)=
\left(\begin{array}{cccc}
-\displaystyle{\frac{1}{{\rm dn}(u_i+u_j)}} & 0 &
0 & \displaystyle{\frac{k {\rm sn}(u_i+u_j)}{{\rm dn}(u_i+u_j)}} \\
0 & -\displaystyle{\frac{{\rm sn}(u_i+u_j)}{{\rm cn}(u_i+u_j)}} &
\displaystyle{\frac{1}{{\rm cn}(u_i+u_j)}} & 0 \\
0 & -\displaystyle{\frac{1}{{\rm cn}(u_i+u_j)}} &
\displaystyle{\frac{{\rm sn}(u_i+u_j)}{{\rm cn}(u_i+u_j)}} & 0 \\
-\displaystyle{\frac{k {\rm sn}(u_i+u_j)}{{\rm dn}(u_i+u_j)}} & 0 &
0 & \displaystyle{\frac{1}{{\rm dn}(u_i+u_j)}}   \end{array} \right)\ ,
\label{e17}
\end{eqnarray}
where ${\rm cn}(u),{\rm dn}(u)$ are also Jacobi's
elliptic functions.  Solutions of tetrahedral Zamolodchikov algebras
are unique in this case, and we obtain the following solutions \
(In ~\cite{Korepanov3}, Korepanov announced that
solutions of tetrahedral Zamolodchikov algebras in the elliptic free
fermion case really satisfy tetrahedron equations, but
did not give explicit expressions of $S^{abc}_{def}$);

\begin{eqnarray}
S^{\rm Sol.}(u_{1},u_{2},u_{3})^{000}_{000}&=&
S^{\rm Sol.}(u_{1},u_{2},u_{3})^{011}_{011}=
S^{\rm Sol.}(u_{1},u_{2},u_{3})^{101}_{101}=
S^{\rm Sol.}(u_{1},u_{2},u_{3})^{110}_{110}=1\ ,
\nonumber\\
S^{\rm Sol.}(u_{1},u_{2},u_{3})^{001}_{010}&=&
\frac{g(u_{2}-u_{3})g(u_{2}+u_{3})}{g(u_{1}-u_{3})g(u_{1}+u_{3})}\ ,
\nonumber\\
S^{\rm Sol.}(u_{1},u_{2},u_{3})^{001}_{100}&=&
(1-k^2)g(u_{2}-u_{3})g(u_{2}+u_{3}) \ ,
\nonumber\\
S^{\rm Sol.}(u_{1},u_{2},u_{3})^{001}_{111}&=&
\frac{1}{g(u_{1}-u_{3})g(u_{1}+u_{3})}\ ,
\nonumber\\
S^{\rm Sol.}(u_{1},u_{2},u_{3})^{010}_{001}&=&
S^{\rm Sol.}(u_{1},u_{2},u_{3})^{010}_{100}=1\ ,
\nonumber\\
S^{\rm Sol.}(u_{1},u_{2},u_{3})^{010}_{111}&=&(1-k^2)\ ,
\label{e20} \\
S^{\rm Sol.}(u_{1},u_{2},u_{3})^{100}_{001}&=&
 -(1-k^2)g(u_{1}-u_{2})g(u_{1}+u_{2}) \ ,
\nonumber\\
S^{\rm Sol.}(u_{1},u_{2},u_{3})^{100}_{010}&=&
\frac{g(u_{1}-u_{2})g(u_{1}+u_{2})}{g(u_{1}-u_{3})g(u_{1}+u_{3})}\ ,
\nonumber\\
S^{\rm Sol.}(u_{1},u_{2},u_{3})^{100}_{111}&=&
-\frac{1}{g(u_{1}-u_{3})g(u_{1}+u_{3})} \ ,
\nonumber\\
S^{\rm Sol.}(u_{1},u_{2},u_{3})^{111}_{001}&=&
g(u_{1}-u_{2})g(u_{1}+u_{2})\ ,
\nonumber\\
S^{\rm Sol.}(u_{1},u_{2},u_{3})^{111}_{010}&=&
-(1-k^2)g(u_{1}-u_{2})g(u_{1}+u_{2})g(u_{2}-u_{3})g(u_{2}+u_{3})\ ,
\nonumber\\
S^{\rm Sol.}(u_{1},u_{2},u_{3})^{111}_{100}&=&
-g(u_{2}-u_{3})g(u_{2}+u_{3})\ ,
\nonumber
\end{eqnarray}
and all other components are zeros, where we use the notation
$\displaystyle{g(u)=\frac{ {\rm sn}(u)} {  {\rm cn}(u) {\rm dn}(u) } }$.

We have explicitly checked that these solutions of
tetrahedral Zamolodchikov algebras really satisfy
tetrahedron equations.

Notice that there is the symmetry
$S^{\rm Sol.}(u_1,u_2,u_3)^{abc}_{def}
=S^{\rm Sol.}(u_3,u_2,u_1)^{cba}_{fed}$, which comes by
taking the transpose of tetrahedral Zamolodchikov algebras.

Under the change of variables from $u_i$ to new variables
$\varphi_i$ in the form
$\tanh(\varphi_i)=(1-k^2)g(u_i)^2$, we can prove the
relations $\tanh(\varphi_i-\varphi_j)
=(1-k^2)g(u_i+u_j)g(u_i-u_j)$.\footnote{We thank Y.G.
Stroganov and J. Hietarinta for
this observation.}  Then we can rewrite $S^{\rm Sol.}$\ as the function
only of the differences of $\varphi_i$.
If we change the transformation of the Boltzmann weights
${\tilde S}^{\rm Sol.} (\varphi_1,\varphi_2,\varphi_3)^{abc}_{def}
=(1-k^2)^{(a+b+c-d-e-f)/2} S^{\rm Sol.}(u_1,u_2,u_3)^{abc}_{def} $,
 which keeps tetrahedron eqs. invariant, we obtain the same expression
as those of Korepanov~\cite{Korepanov2,Korepanov3}
in the following form;
\begin{eqnarray}
{\tilde S}^{\rm Sol.}(\varphi_{1},\varphi_{2},\varphi_{3})^{000}_{000}&=&
{\tilde S}^{\rm Sol.}(\varphi_{1},\varphi_{2},\varphi_{3})^{011}_{011}=
{\tilde S}^{\rm Sol.}(\varphi_{1},\varphi_{2},\varphi_{3})^{101}_{101}=
{\tilde S}^{\rm Sol.}(\varphi_{1},\varphi_{2},\varphi_{3})^{110}_{110}=1\ ,
\nonumber\\
{\tilde S}^{\rm Sol.}(\varphi_{1},\varphi_{2},\varphi_{3})^{001}_{010}&=&
\frac{\tanh(\varphi_{2}-\varphi_{3})}{\tanh(\varphi_{1}-\varphi_{3})}\ ,
\nonumber\\
{\tilde S}^{\rm Sol.}(\varphi_{1},\varphi_{2},\varphi_{3})^{001}_{100}&=&
\tanh(\varphi_{2}-\varphi_{3})\ ,
\nonumber\\
{\tilde S}^{\rm Sol.}(\varphi_{1},\varphi_{2},\varphi_{3})^{001}_{111}&=&
\frac{1}{\tanh(\varphi_{1}-\varphi_{3})}\ ,
\nonumber\\
{\tilde S}^{\rm Sol.}(\varphi_{1},\varphi_{2},\varphi_{3})^{010}_{001}&=&
{\tilde S}^{\rm Sol.}(\varphi_{1},\varphi_{2},\varphi_{3})^{010}_{100}=1\ ,
\nonumber\\
{\tilde S}^{\rm Sol.}(\varphi_{1},\varphi_{2},\varphi_{3})^{010}_{111}&=&1\ ,
\label{e21} \\
{\tilde S}^{\rm Sol.}(\varphi_{1},\varphi_{2},\varphi_{3})^{100}_{001}&=&
 -\tanh(\varphi_{1}-\varphi_{2}) \ ,
\nonumber\\
{\tilde S}^{\rm Sol.}(\varphi_{1},\varphi_{2},\varphi_{3})^{100}_{010}&=&
\frac{\tanh(\varphi_{1}-\varphi_{2})}{\tanh(\varphi_{1}-\varphi_{3})}\ ,
\nonumber\\
{\tilde S}^{\rm Sol.}(\varphi_{1},\varphi_{2},\varphi_{3})^{100}_{111}&=&
-\frac{1}{\tanh(\varphi_{1}-\varphi_{3})} \ ,
\nonumber\\
{\tilde S}^{\rm Sol.}(\varphi_{1},\varphi_{2},\varphi_{3})^{111}_{001}&=&
\tanh(\varphi_{1}-\varphi_{2})\ ,
\nonumber\\
{\tilde S}^{\rm Sol.}(\varphi_{1},\varphi_{2},\varphi_{3})^{111}_{010}&=&
-\tanh(\varphi_{1}-\varphi_{2})\tanh(\varphi_{2}-\varphi_{3}) \ ,
\nonumber\\
{\tilde S}^{\rm Sol.}(\varphi_{1},\varphi_{2},\varphi_{3})^{111}_{100}&=&
-\tanh(\varphi_{2}-\varphi_{3}) \ ,
\nonumber
\end{eqnarray}
Though it is quite non-trivial, we thus obtained the same solutions
of tetrahedron eqs. by using Korepanov mechanism even though
we start from the elliptic case instead of the trigonometric
one in the free fermion model.
Though we cannot find new solutions of tetrahedron eqs. from the elliptic
free fermion model by using Korepanov mechanism, it will be helpful to
understand Korepanov's method and to find
new solutions by using Korepanov mechanism.

Here we give examples which satisfy tetrahedral
Zamolodchikov algebras but not satisfy tetrahedron equations by
considering trigonometric $k=0$\ case.
In this case, tetrahedral Zamolodchikov algebras are not uniquely
solved~\cite{Korepanov2,Korepanov3}, and general solutions can be obtained
by including arbitrary parameters.
We denote
\begin{eqnarray}
S(u_1,u_2,u_3)^{abc}_{def}=S^{\rm Korepa}(u_1,u_2,u_3)^{abc}_{def}+
S^{\rm Homo}(u_1,u_2,u_3)^{abc}_{def}\ ,
\nonumber
\end{eqnarray}
where $S^{\rm Korepa}=S^{\rm Sol.}|_{k=0}$, then $S^{\rm Homo}$\
satisfy homogeneous equations
\begin{eqnarray}
\sum_{d,e,f=0}^{1} S^{\rm Homo}(u_1,u_2,u_3)^{abc}_{def} \
R^f_{23}(u_2,u_3) R^e_{13}(u_1,u_3) R^d_{12}(u_1,u_2)=0 \ ,
\label{e22}
\end{eqnarray}
and we obtain solutions $S^{\rm Homo}(u_1,u_2,u_3)^{abc}_{def}$, which
include arbitrary parameters $h^{\rm even}_{abc}$ and $h^{\rm odd}_{abc}$,
of the following forms
\begin{eqnarray}
&& S^{\rm Homo}(u_1,u_2,u_3)^{abc}_{000}=h^{\rm even}_{abc},
\nonumber \\
&& S^{\rm Homo}(u_1,u_2,u_3)^{abc}_{001}=-h^{\rm odd}_{abc}
\tan(u_1-u_2) \tan(u_1+u_3)\ , \nonumber \\
&& S^{\rm Homo}(u_1,u_2,u_3)^{abc}_{010}=-h^{\rm odd}_{abc}
\tan(u_1-u_2) \tan(u_2-u_3)\ , \nonumber \\
&& S^{\rm Homo}(u_1,u_2,u_3)^{abc}_{011}=h^{\rm even}_{abc}
\cot(u_2-u_3)\cot(u_1+u_3)\ , \nonumber \\
&& S^{\rm Homo}(u_1,u_2,u_3)^{abc}_{100}=h^{\rm odd}_{abc}
\tan(u_2-u_3) \tan(u_1+u_3)\ , \label{e23} \\
&& S^{\rm Homo}(u_1,u_2,u_3)^{abc}_{101}=-h^{\rm even}_{abc}
\cot(u_1-u_2) \cot(u_2-u_3)\ , \nonumber \\
&& S^{\rm Homo}(u_1,u_2,u_3)^{abc}_{110}=-h^{\rm even}_{abc}
\cot(u_1-u_2) \cot(u_1+u_3)\ , \nonumber \\
&& S^{\rm Homo}(u_1,u_2,u_3)^{abc}_{111}=h^{\rm odd}_{abc}\ . \nonumber
\end{eqnarray}
These solutions $S(u_1,u_2,u_3)^{abc}_{def}
=S^{\rm Korepa}(u_1,u_2,u_3)^{abc}_{def}
+S^{\rm Homo}(u_1,u_2,u_3)^{abc}_{def}$
satisfy tetrahedral Zamolodchikov algebras, but not satisfy
tetrahedron equations if some arbitrary
parameres in $h^{\rm even}_{abc}$ and/or $h^{\rm odd}_{abc}$ become
non-zero, that is, tetrahedron equations are satisfied
only the case $S^{\rm Homo}=0$.\vspace{10mm}

%%%%%%%%%%%%%%%%%%%%%%%%%%% section 4 %%%%%%%%%%%%%%%%%%%%%%
\noindent {\bf \S 4. Summary }\vspace{2mm}

\indent
We have examined solutions for tetrahedron
equations from the elliptic free fermion model by
using Korepanov mechanism.
Though it is quite non-trivial, we obtained the same solutions
of tetrahedron eqs. by using Korepanov mechanism even though
we start from the elliptic case instead of the trigonometric
one in the free fermion model. Thus we cannot find new solutions of
tetrahedron eqs. from the elliptic free fermion model by using Korepanov
mechanism, our results will be helpful to understand the Korepanov's
method and to find new solutions by using Korepanov mechanism.
As a byproduct, we have
found a new integrable 2-dim. lattice model, where transfer matrices
are composed of $R^1$\ in Eqs.(\ref{e2}).

We give examples which satisfy tetrahedral Zamolodchikov algebras
but not satisfy tetrahedron equations by solving
tetrahedral Zamolodchikov algebras in trigonometric symmetric free
fermion case. \vspace{10mm}

%\newpage
%%%%%%%%%%%%%%%%%%%Acknowledgement %%%%%%%%%%%%%%%
\noindent {\bf Acknowledgements}\vspace{2mm}

\indent
We would like to thank C. Ahn for helpful discussions
at the early stage of this work, and I.G. Korepanov for
sending us his related papers and his kind help.
One of the authors (K.S.) is grateful to the Special Research
Fund at Tezukayama Univ. for financial support. \vspace{10mm}

%\vfil \break
\newpage
%%%%%%%%%%%%%%%%%%%%%%%%% references %%%%%%%%%%%%%%%%%%%

%%%%%%%%%%%%%%%%%%%%%%%%%%%%%%%%%%%%%%%%%%%%%%%%%%%%%%%%%%%%%%%
\end{document}